\documentclass[copyright,creativecommons]{eptcs}
\providecommand{\event}{ICLP 2026} 

\usepackage{iftex}

\ifpdf
  \usepackage{underscore}         
  \usepackage[T1]{fontenc}        
\else
  \usepackage{breakurl}           
\fi

\usepackage{tikz}
\usetikzlibrary{shapes,positioning,calc,shapes.geometric,arrows.meta,trees}
\usepackage{booktabs}
\usepackage{xspace}
\usepackage{tcolorbox}
\usepackage{listings}
\usepackage[normalem]{ulem}
\definecolor{diffadd}{RGB}{0,128,0}
\definecolor{diffdel}{RGB}{200,0,0}
\lstnewenvironment{codeblock}
  {\lstset{basicstyle=\footnotesize\ttfamily,breaklines=true,
           columns=fullflexible,keepspaces=true,showstringspaces=false,
           escapechar=!,mathescape=false}}
  {}
\newcommand{\ins}[1]{\textcolor{diffadd}{#1}}
\newcommand{\del}[1]{\textcolor{diffdel}{\sout{#1}}}
\lstdefinelanguage{Prolog}{
  morekeywords={is,mod,rem,div,not},
  morecomment=[l]{\%},
  morecomment=[s]{/*}{*/},
  morestring=[b]',
  sensitive=true,
}
\lstdefinestyle{prolog}{
  language=Prolog,
  basicstyle=\footnotesize\ttfamily,
  keywordstyle=\color{c1dark}\bfseries,
  commentstyle=\color{diffdel}\itshape,
  stringstyle=\color{diffadd},
  breaklines=true,
  columns=fullflexible,
  keepspaces=true,
  showstringspaces=false,
}
\usepackage{calc}
\usepackage{multirow}

\definecolor{c1dark}{RGB}{0,114,178}
\definecolor{c2dark}{RGB}{230,159,0}
\definecolor{c1light}{RGB}{214,234,248}
\definecolor{c2light}{RGB}{253,245,207}

\pgfkeyssetvalue{/change colors/none}{red}
\pgfkeyssetvalue{/change colors/RB}{orange}
\pgfkeyssetvalue{/change colors/VMM}{green}

\newcommand{\changed}[2][none]{{\pgfkeysgetvalue{/change colors/#1}{\changecolor}\color{\changecolor}#2}}
\renewcommand{\changed}[2][none]{#2}

\newcommand{\toolname}[0]{\textsc{LogMorph}\xspace}

\newcommand{\circledsm}[1]{\tikz[baseline=-0.8ex]{\node[inner sep=0.1em,circle,fill=gray,text=white] {\sffamily\small #1};}\xspace}

\title{What Bugs Do Prolog Students Write? An Empirical Taxonomy and Data-Driven Mutation Framework}

\author{Ricardo Brancas
\institute{INESC-ID / IST,\\Universidade de Lisboa,\\Portugal}
\and
Pedro Orvalho
\institute{Artificial Intelligence Research Institute,\\
Consejo Superior de Investigaciones Científicas,\\Barcelona, Catalonia, Spain}
\and
Carolina Carreira
\institute{Carnegie Mellon University,\\Pittsburgh, USA}
\institute{INESC-ID / IST, Universidade de Lisboa, Portugal}
\and
Vasco Manquinho
\institute{INESC-ID / IST,\\Universidade de Lisboa,\\Portugal}
\and
Ruben Martins
\institute{Carnegie Mellon University,\\Pittsburgh, USA}}

\def\titlerunning{What Bugs Do Prolog Students Write?}
\def\authorrunning{R. Brancas, P. Orvalho, C. Carreira, V. Manquinho \& R. Martins}
\def\copyrightholders{Brancas, Orvalho, Carreira, Manquinho \& Martins}
\begin{document}
\maketitle

\begin{abstract}
Automated feedback tools for logic programming education depend on realistic bug datasets that reflect the mistakes students actually make.
However, existing mutation testing frameworks for Prolog treat all mutations as equally likely, producing synthetic faults that diverge from classroom reality.
We present an empirical study of 7,201 Prolog submissions from 265 undergraduate students, from which we derive a fine-grained taxonomy of student bugs through manual classification of 200 bug-fixing submissions.
Guided by this taxonomy, we develop \toolname, a data-driven mutation tool whose 17 operators are weighted according to the observed error distribution. \toolname enumerates valid mutation sites on the abstract syntax tree, samples operators proportionally, injects faults, delegating to an SMT-based synthesizer when new code fragments are needed, and validates each mutant against a reference test suite. An evaluation of 16,000 generated mutants shows that the synthetic error distribution closely matches the student distribution, with most bug categories agreeing to within two percentage points.
\changed[RB]{We identify cut-related mutations and synthesizer-generated code as the main sources of residual divergence, and outline how combining the SMT back-end with a language model fine-tuned on student code can further improve realism.}
\end{abstract}

\section{Introduction} \label{sec:introduction}

Teaching logic programming presents a distinctive pedagogical challenge. Unlike imperative or object-oriented paradigms, where students can reason about programs by mentally simulating step-by-step execution, Prolog demands a fundamentally different mode of thinking: declarative specification, unification, and recursive descent through logical clauses. This shift is a well-documented source of difficulty~\cite{duboulay1986,vansomeren1990,taylor1990,brna1990}, and the resulting errors are often different from those in mainstream languages~\cite{fung1990}. Understanding their nature, frequency, and structure is essential for building effective educational tools, yet the empirical landscape of student bugs in logic programming remains underexplored.

Recently, automated feedback tools for programming education have made significant strides~\cite{DBLP:conf/sigcse/OrvalhoJM24,DBLP:conf/pldi/SinghGS13,DBLP:conf/pldi/GulwaniRZ18,DBLP:conf/aaai/GuptaPKS17,DBLP:journals/jss/OrvalhoJM26}, offering students immediate guidance on their submissions. However, the effectiveness of such tools depends critically on how well they anticipate the kinds of mistakes students actually make. A feedback system trained on uniformly distributed synthetic bugs may perform well in aggregate but miss the long-tailed, idiosyncratic errors that dominate real classrooms~\cite{DBLP:conf/sigcse/AlTadmriB15,DBLP:journals/jeric/BrownA17}, a mismatch that is a core obstacle for robust tutoring systems.

Mutation testing~\cite{DBLP:journals/computer/DeMilloLS78,DBLP:journals/tse/Hamlet77}, the systematic injection of small syntactic faults into correct programs, is a natural approach to generating buggy variants for training and evaluation purposes. However, traditional mutation frameworks are designed with software testing in mind: they aim for exhaustive coverage of a predefined set of operators, treating all mutations as equally likely. This assumption is poorly suited to educational contexts, where certain mistake patterns (such as forgetting a base case clause) vastly outnumber others (such as misplacing a cut operator). A more faithful simulation of student behavior requires grounding the mutation process in empirical data about which mistakes students actually make.

In this paper, we address this gap through two complementary contributions. First, we conduct a detailed empirical study of student submissions from a 9-week bachelor's-level logic programming course. We analyze 7201 Prolog code submissions from 265 students across three types of assignments (a role-playing exercise, recitation exercises, and a graded final project) and produce a fine-grained taxonomy of bug types observed in 200 real student programs. Our analysis reveals, among other findings, that incomplete implementations account for the largest share of errors (37.5\%), followed by wrong argument usage (20.5\%) and rule goal problems (13.0\%), with other categories such as operator errors, incorrect predicate names, and cut-related issues occurring at lower frequencies.

Second, we leverage this empirical distribution to build \toolname, a data-driven mutation tool for Prolog. Unlike conventional mutation frameworks, \toolname samples mutation operators according to the observed frequency of each bug type in the student dataset, producing synthetic buggy programs whose error profiles match those of real learners. \toolname operates over a custom Abstract Syntax Tree representation of Prolog programs and implements a four-stage pipeline (enumeration, sampling, injection, and validation) that ensures every generated mutant is both syntactically valid and semantically distinct from the original. For mutations that require generating new code fragments (such as inserting a spurious goal or replacing an argument), \toolname delegates to an SMT-based program synthesizer that produces structurally valid Prolog terms respecting the context of the mutation site.

The contributions of this paper are as follows:

\begin{enumerate}
    \item \textbf{An annotated dataset of student Prolog submissions.} We release a publicly available dataset\footnote{\url{https://figshare.com/s/fca6cb79db0790e85deb}} of 7201 submissions from 265 students, annotated with correctness labels and progression categories (bug fixed, bug introduced, mixed, no change). This dataset provides a foundation for future research on student behavior in logic programming courses.
    \item \textbf{A taxonomy of student bugs in Prolog.} Through manual classification of 200 randomly selected bug-fixing submissions, we identify and categorize the most common error patterns, including subtypes within each major category. This taxonomy offers both quantitative frequency data and concrete examples drawn from real student code.
    \item \textbf{A data-driven mutation tool (\toolname).} We present a mutation framework that uses the empirically derived bug distribution to generate realistic buggy Prolog programs. \toolname's weighted sampling mechanism, combined with a validation loop that rejects trivial or syntactically invalid mutants, produces a synthetic dataset that is representative of actual classroom errors.
    \item \textbf{An empirical evaluation.} We compare the distribution of bug categories in a synthetic dataset\footnotemark[\value{footnote}] of 16,000 mutants against the empirical distribution observed in student submissions, showing close agreement for the majority of categories, and perform a qualitative analysis of the generated mutations, identifying both their strengths and current limitations.
\end{enumerate}

\section{Related Work} \label{sec:related}

\paragraph{Student error taxonomies.}
Empirical classification of novice programming errors has a long history. Pea~\cite{pea1986} identified language-independent conceptual bugs, while Spohrer and Soloway~\cite{DBLP:journals/cacm/SpohrerS86} showed that plan-composition errors dominate in Pascal. For Java, Altadmri and Brown~\cite{DBLP:conf/sigcse/AlTadmriB15} analyzed over 37~million compilations from the Blackbox dataset, and Brown and Altadmri~\cite{DBLP:journals/jeric/BrownA17} showed that educator beliefs about common errors diverge from empirical evidence. Albrecht and Grabowski~\cite{DBLP:conf/sigcse/AlbrechtG20} manually classified 12,371 C~submissions, finding that many errors stem from carelessness rather than misconceptions. In logic programming, the challenges of unification and backtracking are well documented~\cite{vansomeren1990,taylor1990}, and the special issue of \textit{Instructional Science}~\cite{brna1990} advocated structured programming techniques to mitigate ad hoc coding errors. Nevertheless, quantitative bug data remain scarce: Fung et~al.~\cite{fung1990} proposed a qualitative taxonomy of Prolog misconceptions, focusing on execution-model misunderstandings rather than code-level patterns. Our work addresses this gap with a quantitative bug taxonomy grounded in 7,201 real student submissions.

\paragraph{Mutation testing.}
Mutation testing was introduced by DeMillo et~al.~\cite{DBLP:journals/computer/DeMilloLS78} and Hamlet~\cite{DBLP:journals/tse/Hamlet77}, who proposed systematically injecting small faults to evaluate test suites. More recently, Andrews et~al.~\cite{DBLP:conf/icse/AndrewsBL05} and Just et~al.~\cite{DBLP:conf/sigsoft/JustJIEHF14} found significant correlations between mutant and real-fault detection, while Gopinath et~al.~\cite{DBLP:conf/issre/GopinathJG14} showed that fault distributions differ across languages, motivating language-specific operators.
%
For Prolog, Toaldo and Vergilio~\cite{toaldo2006} defined the first mutation operators based on common logic programming mistakes, and Efremidis et~al.~\cite{DBLP:conf/wflp/EfremidisSKK18} built a mutation testing framework for SWI-Prolog and SICStus. Neither Prolog-specific work weighted operators by empirical error frequencies. Our tool derives operator weights from observed classroom data, following the data-driven philosophy of Beller et~al.~\cite{DBLP:conf/icse/BellerWBSM0021}, whose Mutation Monkey at Facebook learns patterns from real bug corpora.

\paragraph{Automated feedback, repair, and debugging.}
Automated feedback for programming education encompasses synthesis-based repair~\cite{DBLP:conf/pldi/SinghGS13,DBLP:conf/pldi/GulwaniRZ18,DBLP:conf/pldi/WangSS18}, semantics-preserving refactoring~\cite{DBLP:conf/kbse/HuAMLR19}, verified repair via MaxSMT~\cite{DBLP:journals/tosem/AhmedFYAR22}, and neural approaches~\cite{DBLP:conf/aaai/GuptaPKS17,DBLP:conf/icse/BhatiaKS18}. All of these target imperative languages. For logic programming, Hong~\cite{DBLP:journals/ijmms/Hong04} proposed grammar-based error analysis for a Prolog tutor, and Brancas et~al.~\cite{DBLP:conf/icst/BrancasMM25} combined logic-based techniques with large language models for Answer Set Programming. On the debugging side, algorithmic debugging of logic programs was pioneered by Shapiro~\cite{shapiro1983} and surveyed by Caballero et~al.~\cite{DBLP:journals/csur/CaballeroRS17}. While such tools diagnose bugs in existing programs, our work instead generates realistic buggy programs, a complementary capability for training and evaluating feedback and debugging methods.

\section{Methodology and Dataset} \label{sec:course}

We evaluated students' behavior during a 9-week-long bachelor's level logic programming course. During this course, students learned classical logic reasoning, as well as logic programming through Prolog. Overall, students faced three different types of Prolog exercises: a 5-part role-playing exercise to familiarize them with the language, 22 optional recitation exercises, and one mandatory, graded final project. The mandatory project was worth 50\% of the final grade, while the remaining 50\% was from other theoretical assessments. Next, we describe the Prolog exercises in more detail.

\begin{itemize}
    \item \textbf{Role-playing Exercise}: Optional, not graded, online only; 5 puzzles over 5 days; a series of simple logic puzzles where students help a detective catch a group of criminals using logic programming.
    \item \textbf{Recitation Exercises}: Optional, not graded, partially solved in-class using pen-and-paper; 22 exercises; 4 Prolog exercise sheets grouped into the topics of: lists, arithmetic, negation, and higher-order functions.
    \item \textbf{Project}: Mandatory, graded, worth 50\% of the final grade; 5 weeks duration; the goal of this year's project was to create a solver for the logic puzzle \textit{star battle}.
\end{itemize}

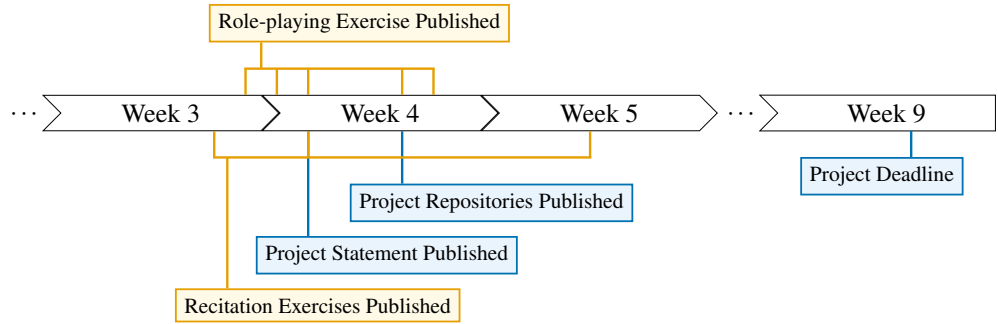
\begin{figure}[tb]
    \centering
    \scalebox{0.9}{\begin{tikzpicture}[
week/.style={signal,draw,signal from=west,minimum width=9em},
c1tag/.style={draw=c1dark,line width=0.75pt,fill=c1light!50!white},
c2tag/.style={draw=c2dark,line width=0.75pt,fill=c2light!50!white},
]
    \node[text height=1.2ex] (e1) {\(\cdots\)};
    \node[week] (w3) [right=0.5em of e1] {Week 3};
    \node[week] (w4) [right=0em of w3] {Week 4};
    \node[week] (w5) [right=0em of w4] {Week 5};
    \node[text height=1.2ex] (e2) [right=0em of w5] {\(\cdots\)};
    \node[week,signal to=nowhere] (w9) [right=0.5em of e2] {Week 9};

    \coordinate (b0312) at ($(w3.south west)!5.5/7!(w3.south east)$);
    \coordinate (t0412) at ($(w3.north west)!6.5/7!(w3.north east)$);
    \coordinate (t0512) at ($(w4.north west)!0.5/7!(w4.north east)$);
    \coordinate (b0612) at ($(w4.south west)!1.5/7!(w4.south east)$);
    \coordinate (t0612) at ($(w4.north west)!1.5/7!(w4.north east)$);
    \coordinate (b0912) at ($(w4.south west)!4.5/7!(w4.south east)$);
    \coordinate (t0912) at ($(w4.north west)!4.5/7!(w4.north east)$);
    \coordinate (t1012) at ($(w4.north west)!5.5/7!(w4.north east)$);
    \coordinate (b1512) at ($(w5.south west)!3.5/7!(w5.south east)$);
    \coordinate (b1301) at ($(w9.south west)!4.5/7!(w9.south east)$);
    
    \draw[line width=1pt,color=c2dark] (b0312) -- +(0,-1em);
    \draw[line width=1pt,color=c2dark] (t0412) -- +(0,1em);
    \draw[line width=1pt,color=c2dark] (t0512) -- +(0,1em);
    \draw[line width=1pt,color=c2dark] (t0612) -- +(0,1em);
    \draw[line width=1pt,color=c1dark] (b0612) -- +(0,-4em);
    \draw[line width=1pt,color=c2dark] (b0612) -- +(0,-1em);
    \draw[line width=1pt,color=c2dark] (t0912) -- +(0,1em);
    \draw[line width=1pt,color=c1dark] (b0912) -- +(0,-2em);
    \draw[line width=1pt,color=c2dark] (t1012) -- +(0,1em);
    \draw[line width=1pt,color=c2dark] (b1512) -- +(0,-1em);
    \draw[line width=1pt,color=c1dark] (b1301) -- +(0,-1em);

    \draw[line width=1pt,color=c2dark] ($(t0412)+(-0.5pt,1em)$) -- ($(t1012)+(0.5pt,1em)$);
    \draw[line width=1pt,color=c2dark] ($(b0312)+(-0.5pt,-1em)$) -- ($(b1512)+(0.5pt,-1em)$);

    \coordinate (b0612end) at ($(b0612)-(0,4em)$);
    \coordinate (b0912end) at ($(b0912)-(0,2em)$);
    \coordinate (b1301end) at ($(b1301)-(0,1em)$);
    
    \node[c1tag,align=center,below right = 0em and -2em,at=(b0612end)] {\footnotesize Project Statement Published};

    \node[c1tag,align=center,below right = 0em and -2em,at=(b0912end)] {\footnotesize Project Repositories Published};

    \node[c1tag,align=center,below left = 0em and -2em,at=(b1301end)] {\footnotesize Project Deadline};

    \draw[line width=1pt,color=c2dark] ($(b0312)+(0.5em,-1em)$) -- ($(b0312)+(0.5em,-6em)$);

    \coordinate (b0312end) at ($(b0312)+(0.5em,-6em)$);

    \node[c2tag,align=center,below right = 0em and -2em,at=(b0312end)] {\footnotesize Recitation Exercises Published};

    \coordinate (t0412mid) at ($(w3.north west)!1!(w3.north east)+(0,1em)$);
    \coordinate (t0412end) at ($(w3.north west)!1!(w3.north east)+(0,2em)$);
    
    \draw[line width=1pt,color=c2dark] (t0412mid) -- (t0412end);

    \node[c2tag,align=center,above right = 0em and -2em,at=(t0412end)] {\footnotesize Role-playing Exercise Published};
    
\end{tikzpicture}}
    \caption{Course exercises timeline.}
    \label{fig:timeline}
\end{figure}

In \autoref{fig:timeline}, we show the timeline of the different exercises solved by the students. During the first few weeks, the students learned logic fundamentals. Then, in week 3, students started learning Prolog, with optional exercise sets released during weeks 3, 4, and 5. Meanwhile, the project statement was released at the beginning of week 4 and lasted until the end of week 9.

\subsection{Dataset Description}

Of the 312 students who made submissions, 265 gave permission for their code to be used for research purposes. Over the course, those 265 students made a total of 7201 code submissions to the different repositories. We have created a publicly available annotated dataset containing these submissions.\footnote{\url{https://figshare.com/s/fca6cb79db0790e85deb}}

\autoref{tab:assignment-performance} shows the total number of correct and incorrect submissions per assignment type, along with the average number of clauses in the students' submissions for each one (per-exercise breakdown is provided in \autoref{tab:assignment-performance-full}). Overall, 1680 submissions passed all tests for the assignment, and 5521 failed at least one test. By far, the assignment with the most number of submissions is the final project, which is easily explained by it being the only mandatory and graded assignment. The ratio of correct to incorrect submissions also varies greatly, with the optional assignments having much fewer incorrect submissions than the project. There are two plausible explanations for this: (1) the optional assignments are easier, and thus take fewer tries to get right, and (2) only more engaged students solved the optional exercises.

\begin{table}[tb]
    \centering
    \small
    \scalebox{0.9}{
    \begin{tabular}{lrrrr}
        \toprule
        \textbf{Assignment} & \textbf{Correct} & \textbf{Incorrect} & \textbf{Total} & \textbf{Avg. Clauses} \\
        \midrule
        {Role-playing Exercise} & 271 & 239 & 510 & 3.51 \\
        {Recitation Exercises} & 969 & 442 & 1411 & 2.50 \\
        {Project} & 440 & 4840 & 5280 & 34.20 \\ \midrule
        \textbf{Total} & 1680 & 5521 & 7201 & 25.81 \\
        \bottomrule
    \end{tabular}
    }
    \caption{Number of correct and incorrect submissions per assignment type. See \autoref{tab:assignment-performance-full} in~\autoref{app:full-table} for the per-exercise breakdown.}
    \label{tab:assignment-performance}
\end{table}

\subsection{Submission Analysis}

\begin{figure}[tb]
    \centering
    \includegraphics{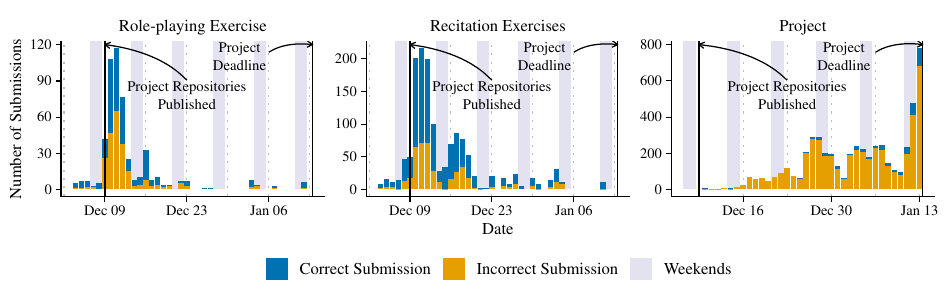}
    \caption{Number of submissions over time for each type of assignment.}
    \label{fig:submissions-time}
\end{figure}

\autoref{fig:submissions-time} shows the number of submissions for each day of the course and reveals interesting patterns in students' behavior. Particularly, most students only start solving the optional exercises on the platform once the project submissions are opened. It is also possible to see the number of submissions escalate over the three days leading up to the project deadline.

\begin{figure}[tb]
    \centering
    \includegraphics{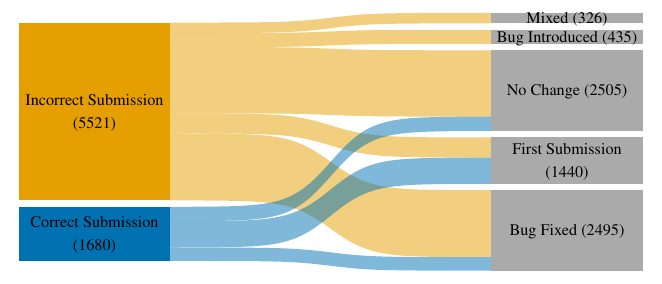}
    \caption{Classification of the 7201 student submissions.}
    \label{fig:submissions-categories}
\end{figure}

We analyzed the 7201 submissions and classified them along two dimensions: (1) whether the submission is correct or incorrect, and (2) whether the submission is an improvement or a regression compared with the previous version of the code. For this second category, we consider 5 different labels:
\begin{itemize}
    \item \textbf{First Submission}: first submission of a student for a particular assignment;
    \item \textbf{Bug Fixed}: the submission passes strictly more tests than the previous version;
    \item \textbf{Bug Introduced}: the submission passes strictly fewer tests than the previous version;
    \item \textbf{Mixed}: the submission both fails tests that previously passed and passes tests that previously failed;
    \item \textbf{No Change}: the submission passes exactly the same tests as the previous version.
\end{itemize}
\autoref{fig:submissions-categories} shows the number of submissions in each category. We observe that most first submissions are correct, but, as evidenced by \autoref{tab:assignment-performance}, this is largely due to the optional exercises. We note that the number of regressions is fairly small (Mixed, Bug Introduced). Lastly, there are a small number of submissions that are purely code quality improvements/refactorings of submissions that were already correct.

\subsection{Bug Analysis}

To understand the types and frequency of bugs, we analyzed submissions under the Bug Fixed label. Of those 2495 submissions, we isolated the ones in which only one predicate was modified -- this subset represents ``ideal'' unit commits in which students fixed a specific incorrect behavior in their program. This selection resulted in 767 submissions, of which we manually classified 200 chosen randomly.

\autoref{tab:bug-types} presents the identified bug types and subtypes found in the 200 selected programs, along with their frequencies and a representative example for each type. Note that a single program may contain multiple bugs, and thus the total number of reported bug types exceeds 200.
The most commonly identified reason for failing submissions was ``Incomplete'' programs, which can happen for two reasons: (1) students implemented their programs incrementally and tested components as they progressed (most often represented by the ``Missing Predicate'' subtype), and (2) students forgot to implement some functionality (most often represented by the ``Missing Clause'' subtype). While the first case does not necessarily represent a true bug, it still gives us a relevant insight: students seem to prefer having many tests to evaluate each functionality in isolation.

\begin{table}[p]
\centering
{\small
\begin{tabular}{p{0.7\textwidth}rr}
    \toprule
    \textbf{Bug Type} & \textbf{Count} & \textbf{Frequency}  \\ \midrule
    \textbf{Incomplete} & \textbf{75} & \textbf{37.5\%} \\
    \small\hspace{2em}\textbf{Subtypes:} Missing Predicate & \small59 & \small29.5\% \\
    \small\hspace{2em}\phantom{\textbf{Subtypes:} }Missing Clause (of existing predicate) & \small16 & \small8.0\% \\
    \hspace*{\fill}\begin{minipage}{\linewidth-2em}
    \small
    There are missing predicates/clauses. Example:\\[-4ex]
    \begin{codeblock}
!\ins{+mult([], \_, []).}!
mult([E1|L1], N, [E2|L2]) :- E2 is E1 * N, mult(L1, N, L2).
\end{codeblock}
\end{minipage} & & \\[-1ex]

\textbf{Wrong Argument} & \textbf{41} & \textbf{20.5\%} \\
\small\hspace{2em}\textbf{Subtypes:} Argument Order Swap & \small7 & \small3.5\% \\
    \small\hspace{2em}\phantom{\textbf{Subtypes:} }Missing Argument & \small6 & \small3.0\% \\
    \small\hspace{2em}\phantom{\textbf{Subtypes:} }Extra Argument & \small2 & \small1.0\% \\
    \small\hspace{2em}\phantom{\textbf{Subtypes:} }Other & \small26 & \small13.0\% \\
\hspace*{\fill}\begin{minipage}{\linewidth-2em}
    \small
    Issue in the arguments of a predicate call. Example:\\[-4ex]
    \begin{codeblock}
mult([], _, []).
mult([E1|L1], N, [E2|L2]) :- E2 is E1 * N, mult(L1, N, !\del{L1}\ins{L2}!).
    \end{codeblock}
    \end{minipage} & & \\[-1ex]
    \textbf{Rule Goal Problems} & \textbf{26} & \textbf{13.0\%} \\
    \small\hspace{2em}\textbf{Subtypes:} Extra Goal & \small13 & \small6.5\% \\
    \small\hspace{2em}\phantom{\textbf{Subtypes:} }Missing Goal & \small10 & \small5.0\% \\
    \small\hspace{2em}\phantom{\textbf{Subtypes:} }Goal Order Swap & \small3 & \small1.5\% \\
    \hspace*{\fill}\begin{minipage}{\linewidth-2em}
    \small
    Problems with the goals in the body of a rule. Example:\\[-4ex]
    \begin{codeblock}
mult([], _, []).
mult([E1|L1], N, [E2|L2]) :- E2 is E1 * N!\ins{, mult(L1, N, L2)}!.
    \end{codeblock}
    \end{minipage} & & \\[-1ex]
    \textbf{Operator Error} & \textbf{23} & \textbf{11.5\%} \\
    \small\hspace{2em}\textbf{Subtypes:} Wrong Operator & \small16 & \small8.0\% \\
    \small\hspace{2em}\phantom{\textbf{Subtypes:} }List Terminators Issue & \small4 & \small2.0\% \\
    \small\hspace{2em}\phantom{\textbf{Subtypes:} }Missing Negation & \small3 & \small1.5\% \\
    \hspace*{\fill}\begin{minipage}{\linewidth-2em}
    \small
    Incorrect operator usage. Example:\\[-4ex]
    \begin{codeblock}
mult([], _, []).
mult([E1|L1], N, [E2|L2]) :- E2 !\del{=}\ins{is}! E1 * N, mult(L1, N, L2).
    \end{codeblock}
    \end{minipage} & & \\[-1ex]
        \textbf{Wrong Variable/Constant} & \textbf{21} & \textbf{10.5\%} \\
    \small\hspace{2em}\textbf{Subtypes:} Wrong Variable Name & \small19 & \small9.5\% \\
    \small\hspace{2em}\phantom{\textbf{Subtypes:} }Wrong Constant & \small2 & \small1.0\% \\
    \hspace*{\fill}\begin{minipage}{\linewidth-2em}
    \small
    Problems with the names of variables or constants. Example:\\[-4ex]
    \begin{codeblock}
mult([], _, []).
mult([E1|L1], N, [E2|L2]) :- E2 is E1 * !\del{M}\ins{N}!, mult(L1, N, L2).
    \end{codeblock}
    \end{minipage} & & \\[-1ex]
    \textbf{Wrong Predicate Name} & \textbf{19} & \textbf{9.5\%} \\
    \hspace*{\fill}\begin{minipage}{\linewidth-2em}
    \small
    Incorrect predicate name in call. Example:\\[-4ex]
    \begin{codeblock}
mult([], _, []).
mult([E1|L1], N, [E2|L2]) :- E2 is E1 * N, mult!\del{iply}!(L1, N, L2).
    \end{codeblock}
    \end{minipage} & & \\[-1ex]
    \textbf{Domain Logic Problem} & \textbf{16} & \textbf{8.0\%} \\
    \hspace*{\fill}\begin{minipage}{\linewidth-2em}
    \small
    Different problems on the implementation of domain logic.
    \end{minipage} & & \\[2ex]
    \textbf{Cut Problem} & \textbf{14} & \textbf{7.0\%} \\
    \small\hspace{2em}\textbf{Subtypes:} Missing Cut & \small10 & \small5.0\% \\
    \small\hspace{2em}\phantom{\textbf{Subtypes:} }Extra Cut & \small3 & \small1.5\% \\
    \small\hspace{2em}\phantom{\textbf{Subtypes:} }Wrong Placement & \small1 & \small0.5\% \\
    \hspace*{\fill}\begin{minipage}{\linewidth-2em}
    \small
    Problems with the cut operator (\texttt{!}). Example:\\[-4ex]
    \begin{codeblock}
max(X, Y, X) :- X >= Y!\ins{, !!}!.
max(_, Y, Y).
    \end{codeblock}
    \end{minipage} & & \\[-1ex]
    \textbf{Other} & \textbf{26} & \textbf{13.0\%} \\
    \bottomrule
\end{tabular}
}
\caption{Description and frequency of the different bug types manually identified in student submissions.} \label{tab:bug-types}
\end{table}

Other bug types were more evenly distributed, with no specific bug occurring at an overwhelmingly high frequency. Of particular note is the ``Domain Logic Problem'' bug type, which reflects a misunderstanding of the intended functionality for a given predicate. Additionally, 13 of the 26 programs categorized as ``Other'' contain at least one predicate that is fundamentally incorrect and requires complete rewriting.

\section{Replicating Student Bugs}

To simulate student behavior and generate a large dataset of buggy Prolog programs, we developed \toolname, a mutation tool. Unlike traditional mutation tools that aim to be exhaustive or statistically uniform, \toolname is \textit{data-driven}: it uses the frequency distribution of bug types identified in \autoref{sec:course} to generate mutations representative of actual student errors. \toolname is implemented in Python and SWI-Prolog using a custom Prolog parser built for this work.

\subsection{Architecture}
The architecture of \toolname is depicted in \autoref{fig:mut-overview}. The pipeline transforms a correct reference implementation into a buggy variant through four stages:

\paragraph{Enumeration.}
The process begins with a \textit{Correct Program} as input. The \textit{Bug Position Collector} parses the source code into an Abstract Syntax Tree (AST) (\circledsm{1}) using our custom parser and traverses the tree to identify all syntactically valid locations where mutation operators can be applied. For instance, the collector identifies every arithmetic operation as a candidate for operator mutation, and every predicate call as a candidate for argument swapping. This step yields a list of \textit{Bug Type/Position Pairs} (\circledsm{2}).

\paragraph{Sampling.}
The \textit{Bug Position Sampler} receives the list of candidate mutations (\circledsm{3}) and selects one according to the weighted probability distribution derived from our manual bug classification (see \autoref{tab:bug-types}). This data-driven weighting ensures that the synthetic dataset mirrors classroom reality; for example, ``Clause Deletion'' mutations are generated significantly more often than cut mutations.

\paragraph{Injection.}
The \textit{Bug Injector} takes the selected pair and rewrites the corresponding AST node to introduce the fault (\circledsm{4}). For operators that require new code (e.g., goal or argument insertion), the injector delegates term generation to a program synthesizer. The synthesizer is SMT-based and built on the Trinity synthesis framework~\cite{DBLP:journals/pvldb/MartinsCCFD19}, following the same approach used by \textsc{ForMHE}~\cite{DBLP:conf/icst/BrancasMM25} for Answer Set Programming. Given the type context of the target position, it generates a syntactically valid Prolog term that satisfies the structural constraints of the surrounding clause.

\paragraph{Validation.}
The mutated program is passed to the \textit{Validator} (\circledsm{5}), which checks that the injected bug produces a \textit{non-trivial} and \textit{observable} fault. Concretely, the validator runs the mutant against the reference test suite and rejects it if it passes all tests (i.e., the mutation is semantically neutral) or fails to load (i.e., the mutation is syntactically invalid). If the mutant is rejected, the validator signals the \textit{Bug Position Sampler} to draw a new candidate (\circledsm{7}), creating a rejection-sampling loop. Once a valid mutant is found, the pipeline emits the final \textit{Buggy Program} (\circledsm{6}).

\begin{figure}[tb]
    \centering
    \scalebox{0.85}{\begin{tikzpicture}[
    very thick,
    thing/.style={draw=c2dark,fill=c2light,inner xsep=0.9em,inner ysep=0.6em,align=center},
    component/.style={draw=c1dark,fill=c1light,inner xsep=0.9em,inner ysep=0.6em,rounded corners=0.5em,align=center},
    numcircle/.style={fill=gray,inner sep=0.15em,circle,font=\sffamily\bfseries\color{white}\small}
]

    \node[thing] (start) {Correct\\Program}; 
    
    \node[component] (enum) [right=5em of start] {Bug Position\\Collector};
    
    \node[thing] (pairs) [right=5em of enum] {Bug Type/Position\\Pairs};
    
    \node[component] (sampler) [below=3.5em of pairs] {Bug Position\\Sampler};
    
    \node[component] (injector) [left=3.5em of sampler] {Bug\\Injector};
    
    \node[component] (validator) [left=3.5em of injector] {Validator};
    
    \node[thing] (end) [left=3.5em of validator] {Buggy\\Program};

    
    \draw[arrows = {-Straight Barb[length=0.4em]},thick,rounded corners=4pt] 
        (start) -- node [numcircle,above=2pt] {1} (enum);

    \draw[arrows = {-Straight Barb[length=0.4em]},thick,rounded corners=4pt] 
        (enum) -- node [numcircle,above=2pt] {2} (pairs);

    \draw[arrows = {-Straight Barb[length=0.4em]},thick,rounded corners=4pt] 
        (pairs) -- node [numcircle,right=2pt] {3} (sampler);

    \draw[arrows = {-Straight Barb[length=0.4em]},thick,rounded corners=4pt] 
        (sampler) to node [numcircle,above=2pt] {4} (injector);


    \draw[arrows = {-Straight Barb[length=0.4em]},thick,rounded corners=4pt] 
        (injector) -- node [numcircle,above=2pt] {5} (validator);

    \draw[arrows = {-Straight Barb[length=0.4em]},thick,rounded corners=4pt] 
        (validator) -- node [numcircle,above=2pt] {6} (end);

    \draw[arrows = {-Straight Barb[length=0.4em]},thick,rounded corners=4pt] 
        (validator) to [bend left=32] node [numcircle,above=2pt] {7} (sampler);

\end{tikzpicture}}
    \caption{Architecture of \toolname.} \label{fig:mut-overview}
\end{figure}

\subsection{Mutation Operators}

We implemented a set of mutation operators that map directly to the bug categories identified in \autoref{sec:course}, with sampling weights proportional to the observed frequencies in \autoref{tab:bug-types}. One deliberate exception is the \textit{Missing Predicate} subtype of ``Incomplete'' bugs: as discussed in \autoref{sec:course}, these submissions typically reflect students testing partial implementations incrementally rather than genuine programming errors, and are therefore excluded. After removing the 59 submissions whose only bug was a missing predicate, 141 programs remain as the base for computing relative frequencies. The denominator in each weight is not always 141 because not all bug types are applicable to all programs; for instance, only 31 of the 141 programs use negation, so the \textit{Missing Negation} frequency is computed over those 31 programs.

For easier comparison with the student data, fine-grained mutation operations are grouped into broader shared categories: ``Domain Logic Problem'', ``Missing Predicate Call'', ``Missing Arithmetic Evaluation'', ``Goal Failure Suppression'', and ``Extraneous Code'' are all mapped to the \textit{Other} category, while ``Missing List Terminators'' and ``Extra List Terminators'' are merged into \textit{List Terminators Issue}. \autoref{tab:operators} lists all implemented operators together with their weights and the shared category label.

Two weights include a manual correction to compensate for a systematic bias introduced by the validation step. Because the sampler and validator operate in a rejection-sampling loop, operators with a high validation pass rate are effectively over-sampled in the final output, while operators that are frequently rejected are under-sampled. Concretely, the \textit{Variable Replacement} weight is reduced by 6 percentage points, and the \textit{Argument Mutation} weight by 4 percentage points relative to their observed frequencies, since variable and argument replacements almost invariably produce detectable failures and therefore pass validation at a near-100\% rate.

\begin{table}[tb]
    \centering
    \footnotesize
    \begin{tabular}{@{}lllrr@{}}
        \toprule
        \textbf{Category} & \textbf{Operator} & \textbf{Description} & \textbf{Rel.\ Freq.} & \textbf{Weight} \\
        \midrule
        Incomplete & Clause Deletion & Remove a clause & $16/141$ & 11.35 \\
        Wrong Var.\ Name & Variable Repl. & Replace a variable in scope & $19/141^{\dagger}$ & 7.48 \\
        Wrong Constant & Constant Repl. & Replace a constant & $2/99$ & 2.02 \\
        Wrong Argument & Argument Mutation & Replace an argument & $26/141^{\ddagger}$ & 14.44 \\
        Missing Argument & Argument Removal & Remove an argument & $6/141$ & 4.26 \\
        Extra Argument & Argument Addition & Add an argument & $2/141$ & 1.42 \\
        Swap Argument & Argument Swap & Swap two arguments & $7/136$ & 5.15 \\
        Extra Goal & Goal Insertion & Insert a sub-goal & $13/141$ & 9.22 \\
        Rule Goal Swap & Goal Swap & Reorder sub-goals & $3/134$ & 2.24 \\
        Missing Goal & Goal Deletion & Remove a sub-goal & $10/141$ & 7.09 \\
        Wrong Operator & Operator Mutation & Change an operator & $16/120$ & 13.33 \\
        Missing Negation & Negation Removal & Remove \texttt{\textbackslash+} & $3/31$ & 9.68 \\
        \multirow{2}{*}{List Term.\ Issue} & Missing List Term. & Remove list brackets & $3/129$ & 2.33 \\
         & Extra List Term. & Add list brackets & $1/141$ & 0.71 \\
        Missing Cut & Cut Removal & Remove a cut (\texttt{!}) & $10/79$ & 12.66 \\
        Extra Cut & Cut Addition & Add a cut (\texttt{!}) & $3/141$ & 2.13 \\
        Wrong Pred.\ Name & Predicate Rename & Replace a predicate name & $19/141$ & 13.48 \\
        \midrule
        \multirow{5}{*}{Other} & Missing Pred.\ Call & Remove a predicate call & $1/141$ & 0.71 \\
         & Missing Arith.\ Eval. & Remove \texttt{is} evaluation & $1/89$ & 1.12 \\
         & Goal Failure Suppr. & Suppress goal failure & $5/27$ & 18.52 \\
         & Extraneous Code & Insert extraneous code & $1/141$ & 0.71 \\
         & Code Injection/Repl. & Replace an AST node & $18/141$ & 12.76 \\
        \bottomrule
        \multicolumn{5}{@{}l}{$^{\dagger}$\,Bias: $-0.06$ \quad $^{\ddagger}$\,Bias: $-0.04$. Weight includes bias.} \\
    \end{tabular}
    \caption{Mutation operators, descriptions, and sampling weights. The \textit{Category} column shows the shared label used for comparison with student data in \autoref{fig:bugcats}. Operators marked with $^{\dagger}$ and $^{\ddagger}$ include a manual bias correction (see text).}
    \label{tab:operators}
\end{table}

\subsection{Worked Example}

We trace the pipeline on a simple \texttt{member/2} predicate to make each stage concrete.

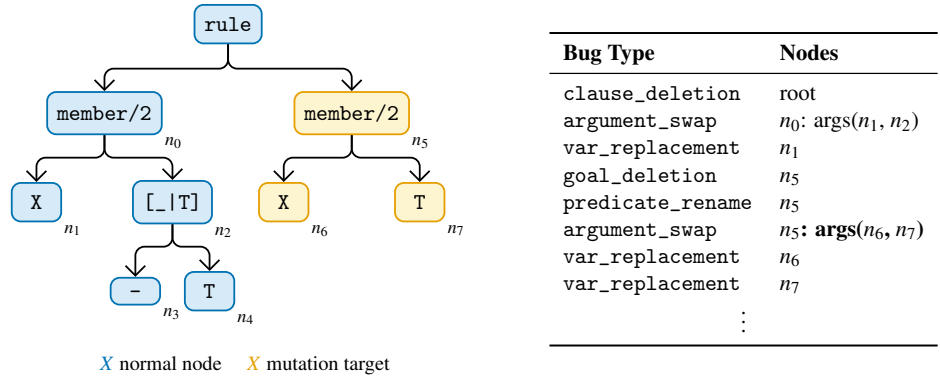
\begin{figure}[tb]
    \centering
    \scalebox{0.85}{
    \begin{minipage}[c]{0.54\linewidth}
        \centering
        \begin{tikzpicture}[
            thick,
            level distance=8ex,
            level 1/.style={sibling distance=10em},
            level 2/.style={sibling distance=5.5em},
            level 3/.style={sibling distance=3em},
            level/.style={arrows={-Straight Barb[length=0.4em]},thick},
            tnode/.style={rounded corners=4pt,inner xsep=0.4em,inner ysep=0.5em,
                          font=\ttfamily\small,minimum width=2em},
            edge from parent/.style={draw,rounded corners=4pt},
            filled/.style={tnode,draw=c1dark,fill=c1light},
            target/.style={tnode,draw=c2dark,fill=c2light},
            edge from parent fork down]

            \node[filled] (root) {rule}
                child {node[filled] (n0) {member/2}
                    child {node[filled] (n1) {X}}
                    child {node[filled] (n2) {[\_|T]}
                        child {node[filled] (n3) {\_}}
                        child {node[filled] (n4) {T}}
                    }
                }
                child {node[target] (n5) {member/2}
                    child {node[target] (n6) {X}}
                    child {node[target] (n7) {T}}
                };

            \begin{scope}[nodes={below right=-3pt and -3pt},font=\scriptsize]
                \foreach \x in {0,...,7}{\node at (n\x.south east) {$n_{\x}$};}
            \end{scope}
        \end{tikzpicture}
        \smallskip

        {\footnotesize\textcolor{c1dark}{$X$}~normal node \quad
                       \textcolor{c2dark}{$X$}~mutation target}
    \end{minipage}%
    \begin{minipage}[c]{0.43\linewidth}
        \centering
        \small
        \begin{tabular}{ll}
            \toprule
            \textbf{Bug Type} & \textbf{Nodes} \\
            \midrule
            \texttt{clause\_deletion}  & root \\
            \texttt{argument\_swap}    & $n_0$: args($n_1$, $n_2$) \\
            \texttt{var\_replacement}  & $n_1$ \\
            \texttt{goal\_deletion}    & $n_5$ \\
            \texttt{predicate\_rename} & $n_5$ \\
            \textbf{\texttt{argument\_swap}} & $n_5$\textbf{: args(}$n_6$\textbf{,}~$n_7$\textbf{)} \\
            \texttt{var\_replacement}  & $n_6$ \\
            \texttt{var\_replacement}  & $n_7$ \\
            \multicolumn{2}{c}{$\vdots$} \\
            \bottomrule
        \end{tabular}
    \end{minipage}
    }
    \caption{AST for clause~2 of \texttt{member/2} (left) and a subset of the bug type/position pairs (right). The bold row is the pair selected by the sampler; highlighted nodes ($n_5$--$n_7$) are the mutation target.}
    \label{fig:ast-example}
\end{figure}

\paragraph{Input.}
The input is the following correct two-clause definition:
\begin{lstlisting}[style=prolog]
member(X, [X|_]).
member(X, [_|T]) :- member(X, T).
\end{lstlisting}

\paragraph{Enumeration.}
The parser reads the source and builds an AST for each clause. \autoref{fig:ast-example} shows the AST for clause~2. The root \texttt{rule} node has one child per element of the clause: the head ($n_0$) and each body goal (here only $n_5$). Each functor node carries its arguments as children, and list patterns such as~\texttt{[\_|T]} are represented as a two-child node ($n_2$) with an anonymous variable ($n_3$) and the tail variable ($n_4$). The \textit{Bug Position Collector} traverses the AST and emits one pair for every applicable operator at every applicable node. A representative subset for \texttt{member/2} is listed in \autoref{fig:ast-example}.

\paragraph{Sampling.}
The sampler draws from the candidate list using the weighted distribution calibrated to our bug data. Suppose it selects the bold row in \autoref{fig:ast-example}:
\(
  \langle\, \texttt{argument\_swap},\ n_5\texttt{: args}(n_6,\, n_7) \,\rangle
\)

\paragraph{Injection.}
The injector locates node~$n_5$ in the AST and swaps its argument children $n_6$ and~$n_7$, turning \texttt{member(X,\,T)} into \texttt{member(T,\,X)}, yielding:
\begin{lstlisting}[style=prolog]
member(X, [X|_]).
member(X, [_|T]) :- member(T, X).  % args swapped
\end{lstlisting}

\paragraph{Validation.}
The mutant is run against the reference test suite. 
\texttt{member(a,[a,b,c])} is still handled by the base clause, but the recursive call now diverges on \texttt{member(b,[a,b,c])}, so the mutant correctly fails tests and is accepted by the validator.

\section{Analysis}

We evaluate \toolname along two dimensions: how well the synthetic bug distribution matches real student errors, and how realistic the generated mutations are in practice.

\subsection{Distribution Comparison}

\begin{figure}
    \centering
    \includegraphics{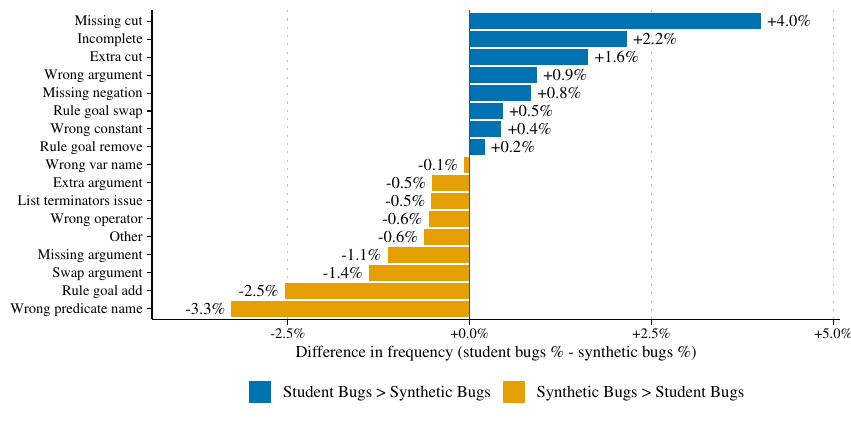}
    \caption{Distribution of bug categories in the student dataset (population) and in the 16,000 synthetically generated mutations (synthetic), as a percentage of each total.} \label{fig:bugcats}
\end{figure}

We applied \toolname to the reference solutions of the student exercises to generate a synthetic dataset of 16,000 buggy programs. \autoref{fig:bugcats} compares the resulting distribution of bug categories against the empirical distribution observed in the student dataset.

Overall, the two distributions are in close agreement. For most categories, the difference between the population and synthetic percentages is less than 2 percentage points, confirming that the data-driven weighting mechanism successfully replicates the overall shape of real student error patterns. The three most frequent categories, \textit{Other} (20.79\% vs.\ 21.42\%), \textit{Wrong argument} (15.35\% vs.\ 14.43\%), and \textit{Wrong predicate name} (9.41\% vs.\ 12.68\%), account for roughly 45\% of both the real and synthetic bugs.

The most pronounced discrepancy concerns cut-related bugs. In the student dataset, \textit{Missing cut} accounts for 4.95\% of bugs, and \textit{Extra cut} for a further 1.98\%, for a combined total of 6.93\%. In the synthetic data, these categories shrink to 0.95\% and 0.35\%, respectively (combined: 1.30\%). This under-representation is a direct consequence of the validation step: the rejection-sampling loop discards mutations that leave the test outcomes unchanged, and cut insertions or deletions are disproportionately rejected because many student predicates either do not use cuts or are not sensitive to them under the available tests.

In the opposite direction, \textit{Wrong predicate name} is overrepresented in the synthetic data (12.68\%) compared with the population (9.41\%), a difference of 3.28 percentage points. \textit{Rule goal add} shows a similar pattern (7.49\% vs.\ 4.95\%), likely because synthesized goals that do alter program semantics pass validation at a higher-than-average rate.

\subsection{Qualitative Analysis and Limitations}

We also examined individual mutations to assess their qualitative realism. For most operator types, the generated bugs are concise, local changes that closely resemble actual student errors. Consider the following three examples drawn from the synthetic dataset:

\begin{enumerate}
    \item \textbf{Wrong operator.} In a list-multiplication predicate, the arithmetic evaluation operator \texttt{is} is replaced with the unification operator \texttt{=}:
\begin{codeblock}
multPorN([], _, []).
multPorN([H | T], N, [H1 | T1]) :- H1 !\del{is}\ins{=}! H * N, multPorN(T, N, T1).
\end{codeblock}

    \item \textbf{Wrong variable name.} In a list-duplication predicate, the tail variable \texttt{T} in the recursive call is replaced with the accumulator variable \texttt{X}:
\begin{codeblock}
duplicaElem([], []).
duplicaElem([H | T], L) :- duplicaElem(!\del{T}\ins{X}!, X), L = [H, H | X].
\end{codeblock}

    \item \textbf{Missing recursive call.} The recursive goal is deleted from a clause body, turning the rule into a predicate that processes only the first element:
\begin{codeblock}
multPorN([], _, []).
multPorN([H | T], N, [W | WT]) :- W is H * N!\del{, multPorN(T, N, WT)}!.
\end{codeblock}
\end{enumerate}

However, mutations that rely on the SMT-based synthesizer to generate new code fragments (specifically goal insertion, argument addition, and extraneous code injection) can produce visibly artificial results. Two representative examples illustrate this:

\begin{enumerate}
    \item \textbf{Synthesized guard on a fact.} A goal insertion operator turns a base-case fact into a rule by adding the synthesized goal \texttt{[] < V0}:
\begin{lstlisting}[style=prolog]
eliminaNumeros([], []) :- [] < V0.   % guard added
\end{lstlisting}
    Comparing an empty list with a fresh variable using \texttt{<} is syntactically valid but semantically nonsensical; no student would write such a guard.

    \item \textbf{Extraneous synthesized clause.} An extraneous code operator injects an entire fabricated clause between two existing ones:
\begin{lstlisting}[style=prolog]
mult(N1, N2, N3) :- N3 is N1 * N2.
multPorN(_, _, _) :- [] * false = _,             % injected
                     multPorN(_, V0, _) =:= (\+[]).
multPorN(P1, N, P2) :- maplist(mult(N), P1, P2).
\end{lstlisting}
    The injected clause exhibits typical traits of SMT synthesis: fresh variables (\texttt{V0}), Boolean constants (\texttt{false}), empty lists used as arithmetic operands, and operators applied to semantically incompatible arguments.
\end{enumerate}

These artifacts arise because the Trinity-based synthesizer~\cite{DBLP:journals/pvldb/MartinsCCFD19} explores the space of structurally valid terms without any model of idiomatic Prolog usage. Since Prolog is untyped, expressions such as \texttt{[] < V0} or \texttt{[] * false} are syntactically legal and accepted by the interpreter, so the validation step, which only checks whether the mutant produces different test outcomes, cannot reject them. The result is code that, while technically valid and behavior-altering, is immediately recognizable as artificial.

Replacing the SMT synthesizer with a large language model fine-tuned on student code would address this limitation: an LLM could generate insertions and replacements that respect both structural constraints and stylistic conventions, producing more realistic mutations for the operator types that currently require synthesis.

\section{Conclusion}

We have presented an empirical study of student errors in Prolog and \toolname, a data-driven mutation tool that uses the resulting bug taxonomy to generate realistic buggy programs. Our evaluation shows that \toolname's weighted sampling mechanism produces synthetic datasets whose distribution closely matches that of real student errors, with most bug categories differing by fewer than 2 percentage points.

The main open direction is to explore the use of large language models to complement the current SMT-based synthesizer, potentially enabling the generation of code fragments that better adhere to both structural and stylistic conventions. We also plan to integrate \toolname into automated feedback systems for logic programming and to test the approach with other logic programming languages and larger, multi-institutional student populations.

\section*{Acknowledgments}
This work supported by Portuguese national funds through FCT, under project 2023.14280.PEX (DOI: 10.54499/\-2023.14280.PEX). This work was also supported by grant PID2022-139835NB-C21 funded by MCIN/\-AEI/\-10.13039/\-501100011033 and by ERDF, EU; and by HORIZON-MSCA-2025-PF, project 101269051 — Sherlock4Py funded by REA, EU. This work was additionally supported by the Carnegie Mellon University Portugal Program and FCT under grants PRT/BD/152086/2021 (DOI: 10.54499/\-PRT/\-BD/\-152086/\-2021) and PRT/BD/153739/2021 (DOI: 10.54499/\-PRT/\-BD/\-153739/\-2021). This work was also partially supported by the National Science Foundation (NSF) under Award CCF2427581 and DARPA under Agreement FA8750-24-9-1000.

\bibliographystyle{eptcs}
\bibliography{generic}

\clearpage

\appendix

\section{Per-Exercise Submission Breakdown}
\label{app:full-table}

\begin{table}[h]
    \centering
    \small
    \begin{tabular}{lrrrr}
        \toprule
        \textbf{Assignment} & \textbf{Correct} & \textbf{Incorrect} & \textbf{Total} & \textbf{Avg. Clauses} \\
        \midrule
        \textbf{Role-playing Exercise} & 271 & 239 & 510 & 3.51 \\
        \hspace{1.5em}Day 1 & 88 & 58 & 146 & 3.00 \\
        \hspace{1.5em}Day 2 & 63 & 67 & 130 & 3.27 \\
        \hspace{1.5em}Day 3 & 40 & 68 & 108 & 5.05 \\
        \hspace{1.5em}Day 4 & 46 & 9 & 55 & 2.62 \\
        \hspace{1.5em}Day 5 & 34 & 37 & 71 & 3.26 \\
        \textbf{Recitation Exercises} & 969 & 442 & 1411 & 2.50 \\
        \hspace{1.5em}Exercise A1 & 65 & 55 & 120 & 2.00 \\
        \hspace{1.5em}Exercise A2 & 79 & 44 & 123 & 2.67 \\
        \hspace{1.5em}Exercise A3 & 61 & 24 & 85 & 2.05 \\
        \hspace{1.5em}Exercise A4 & 49 & 15 & 64 & 2.00 \\
        \hspace{1.5em}Exercise A5 & 49 & 12 & 61 & 3.05 \\
        \hspace{1.5em}Exercise A6 & 50 & 21 & 71 & 2.00 \\
        \hspace{1.5em}Exercise A7 & 46 & 19 & 65 & 2.08 \\
        \hspace{1.5em}Exercise B1 & 57 & 42 & 99 & 3.09 \\
        \hspace{1.5em}Exercise B2 & 39 & 31 & 70 & 1.77 \\
        \hspace{1.5em}Exercise B3 & 42 & 16 & 58 & 3.55 \\
        \hspace{1.5em}Exercise B4 & 41 & 7 & 48  & 1.70 \\
        \hspace{1.5em}Exercise B5 & 45 & 15 & 60 & 2.86 \\
        \hspace{1.5em}Exercise C1 & 41 & 19 & 60 & 2.22 \\
        \hspace{1.5em}Exercise C2 & 36 & 6 & 42 & 2.98 \\
        \hspace{1.5em}Exercise C3 & 35 & 7 & 42 & 3.32 \\
        \hspace{1.5em}Exercise C4 & 37 & 17 & 54 & 3.11 \\
        \hspace{1.5em}Exercise D1 & 47 & 13 & 60 & 2.25 \\
        \hspace{1.5em}Exercise D2 & 35 & 10 & 45 & 2.56 \\
        \hspace{1.5em}Exercise D3 & 36 & 14 & 50 & 3.09 \\
        \hspace{1.5em}Exercise D4 & 33 & 23 & 56 & 3.02 \\
        \hspace{1.5em}Exercise D5 & 23 & 27 & 50 & 2.12 \\
        \hspace{1.5em}Exercise D6 & 23 & 5 & 28 & 2.00 \\
        \textbf{Project} & 440 & 4840 & 5280 & 34.20 \\ \midrule
        \textbf{Total} & 1680 & 5521 & 7201 & 25.81 \\
        \bottomrule
    \end{tabular}
    \caption{Number of correct and incorrect submissions per exercise.}
    \label{tab:assignment-performance-full}
\end{table}

\end{document}